\documentclass[aps,prl,reprint,floatfix,amsmath,boldmath]{revtex4-1} % PRL preprint class
\usepackage[pdftex]{color} %the build LaTeX => PDF
\usepackage{hyphenat}
\usepackage[pdftex,colorlinks=true,allcolors=blue,breaklinks=true]{hyperref}  %the build LaTeX => PDF
\usepackage[normalem]{ulem} 
\usepackage{soul}

\definecolor{g-blue}{rgb}{0.83,0.95,1}
\definecolor{g-yellow}{rgb}{1,1,0.7}
\definecolor{g-green}{rgb}{0.9,1,0.9}
\definecolor{green}{rgb}{0,0.6,0}
\definecolor{cyan}{rgb}{0,0.7,0.7}
\definecolor{black}{rgb}{0,0,0}
\definecolor{grey}{rgb}{0.4 ,0.4 ,0.4 }

\usepackage{bm}
\usepackage{amssymb}
\usepackage{amsfonts}
\usepackage{amsmath}
\usepackage{graphicx}
\usepackage{amsmath,bm,epsfig}

 \def \ed {\end{document}}
\def\Fbox#1{\vskip1ex\hbox to 8.5cm{\hfil\fboxsep0.3cm\fbox{%
		\parbox{8.0cm}{#1}}\hfil}\vskip1ex\noindent}  %%  {TEXT} in BOX
		 
\newcommand{\Fig}[1]{Fig.\,\ref{#1}}%%  requires \Fef{label}
\newcommand{\Ref}[1]{Ref.\,\cite{#1}}%%  requires \Fef{label}
\newcommand{\BE}[1]{\begin{equation}\label{#1}}

\def\be{\begin{equation}}\def\ee{\end{equation}}
\def\bea{\begin{eqnarray}}\def\eea{\end{eqnarray}}
\def\bse{\begin{subequations}}\def\ese{\end{subequations}}

\let \= \equiv \let\*\cdot \let\~\widetilde \let\^\widehat \let\-\overline

 \def\1{\bm1} 

\def\<{\left\langle}    \def\>{\right\rangle}
\def\({\left(}          \def\){\right)}
\def \[ {\left [} \def \] {\right ]}

\newcommand{\B}[1]{{\bm{#1}}}%% Bold Roman & Greek Lower & Upper Case
\renewcommand{\sb}[1]{_{\text {#1}}}  %% sub-   for lower case
\def\Sb#1{_{\scriptscriptstyle\rm{#1}}}

\begin{document}
	
\title{Tunable space-time crystal in room-temperature magnetodielectrics}

\author{Alexander~J.~E.~Kreil}	
\email{kreil@rhrk.uni-kl.de}
\affiliation{Fachbereich Physik and Landesforschungszentrum OPTIMAS, Technische Universit\"at Kaiserslautern, 67663 Kaiserslautern, Germany \looseness=-1}

\author{Halyna~Yu.~Musiienko-Shmarova}
\affiliation{Fachbereich Physik and Landesforschungszentrum OPTIMAS, Technische Universit\"at Kaiserslautern, 67663 Kaiserslautern, Germany \looseness=-1}

\author{Dmytro~A.~Bozhko}
\affiliation{Fachbereich Physik and Landesforschungszentrum OPTIMAS, Technische Universit\"at Kaiserslautern, 67663 Kaiserslautern, Germany \looseness=-1}

\author{Anna~Pomyalov}
\affiliation{Department of Chemical and Biological Physics, Weizmann Institute of Science, Rehovot 76100, Israel \looseness=-1}

\author{Victor~S.~L'vov}
\affiliation{Department of Chemical and Biological Physics, Weizmann Institute of Science, Rehovot 76100, Israel \looseness=-1}

\author{Sebastian~Eggert}
\affiliation{Fachbereich Physik and Landesforschungszentrum OPTIMAS, Technische Universit\"at Kaiserslautern, 67663 Kaiserslautern, Germany \looseness=-1}

\author{Alexander~A.~Serga}
\affiliation{Fachbereich Physik and Landesforschungszentrum OPTIMAS, Technische Universit\"at Kaiserslautern, 67663 Kaiserslautern, Germany \looseness=-1}

\author{Burkard~Hillebrands}
\affiliation{Fachbereich Physik and Landesforschungszentrum OPTIMAS, Technische Universit\"at Kaiserslautern, 67663 Kaiserslautern, Germany \looseness=-1}

\begin{abstract}
We report the experimental realization of a space-time crystal with tunable periodicity in time and space in the magnon Bose-Einstein Condensate (BEC), formed in a room-temperature Yttrium Iron Garnet (YIG) film by radio-frequency space-homogeneous magnetic field. The magnon BEC is prepared to have a well defined frequency and non-zero wavevector. We demonstrate how the crystalline ``density" as well as the time and space textures of the resulting crystal may be tuned by varying the experimental parameters: external static magnetic field, temperature, thickness of the YIG film and power of the radio-frequency field. The proposed space-time crystals provide a new dimension for exploring dynamical phases of matter and  can serve as a model nonlinear Floquet system, that brings in touch the rich fields of classical nonlinear waves, magnonics and periodically driven systems. \looseness=-1
\end{abstract}

\maketitle 

Spontaneous symmetry breaking is a fundamental concept of physics. A well known example is the breaking of spatial translational symmetry, which leads to a phase transition from fluids to solid crystals. By analogy, one can think about a ``time crystal" as the result of breaking translational symmetry in time. More generally, one expects the appearance of a ``space-time crystal" as a consequence of breaking translational symmetry both in time and in space. If a time crystal exists it  should demonstrate time-periodic motion of its ground state \cite{1}. In addition to the time periodicity, the space-time crystals should be periodic in space, similar to an ordinary crystal. \looseness=-1

It was recently  argued that  time- and space-time crystals cannot be realized in thermodynamic equilibrium\,\cite{2,3}. This led to a search of space-time symmetry breaking in a wider context, for example in a system with flux-equilibrium, rather than in the thermodynamic equilibrium. Needless to say that oscillatory non-equilibrium states are well known already.  One can remember, for example, gravity waves exited on a sea surface under windy conditions, quasi-periodic current instabilities in some semiconductors carrying strong DC electric current (a Gunn effect\,\cite{Gunn} in A$_3$B$_5$ semiconductors, such as $n$-type GaAs, used in police radar speed guns), instabilities of an electron beam in plasma. More recent similar example is microwave generation by nano-sized magnetic oscillators driven by a spin-polarized DC electric current \cite{nanooscillators}. On the other hand new physical phenomena can be observed in time-crystals which do not absorb or dissipate energy from external pumping, but instead build up a {\it coherent} time-periodic quantum state due to the complexity in the interactions. Possibly a first nontrivial example, found in Ref.\,\cite{4}, was a periodically driven Floquet quantum disordered system, demonstrating subharmonic behavior. Non trivially, this system does not actually absorb and dissipate any of the externally pumped energy because the disorder in the system makes the energy states isolated from one another, see also Refs.\,\cite{8,9}. The possibly most recent observation of a time crystal with very long energy relaxation time compared to energy-conserving interaction processes is the Bose-Einstein condensate (BEC) of magnons in a flexible trap in superfluid $^3$He-\textit{B} under periodic driving by an external magnetic field \cite{volovik}. Other important requirements for the existence of time crystals include the robustness---the independence of their features to the perturbations of the physical system (e.g. level of disorder) and appearance of the soft modes \cite{5,6,7}. The problems related to space-time crystals \cite{STC-Exp} are more involved and up to now were explored mostly theoretically\,\cite{STC-1,STC-2,STC-3,STC-4,STC-5}. 
 
In this paper we report the experimental realization of a space-time crystal with tunable periodicity in time and space in the magnon BEC \cite{Demokritov2006,Nowik-Boltyk2012,Giamarchi2008,Bozhko2016,Kreil2018,Serga2014}, formed in a room-temperature Yttrium Iron Garnet (YIG, $\mathrm{Y_3Fe_5O_{12}}$) film. The condensate spontaneously arises as a result of scattering of parametrically injected magnons to the bottom of their spectrum. The scattered magnons initially form spectrally localized groups, which can be best described as a time-polycrystal with partial coherence. After switching off the pumping, we observe an interaction-driven condensation into two coherent spatially extended spin waves---magnon BECs---which are best characterized by a space-time crystal. We show that this coherent state has the hallmark of non-universal relaxation times, which are much longer than the intrinsic time scales and the crystallization time. We consider the magnon BEC as a model object in studies of Floquet nonlinear wave systems, subject to intensive periodical in time impact. 

\begin{figure}[t]
	\includegraphics[width=0.98\columnwidth]{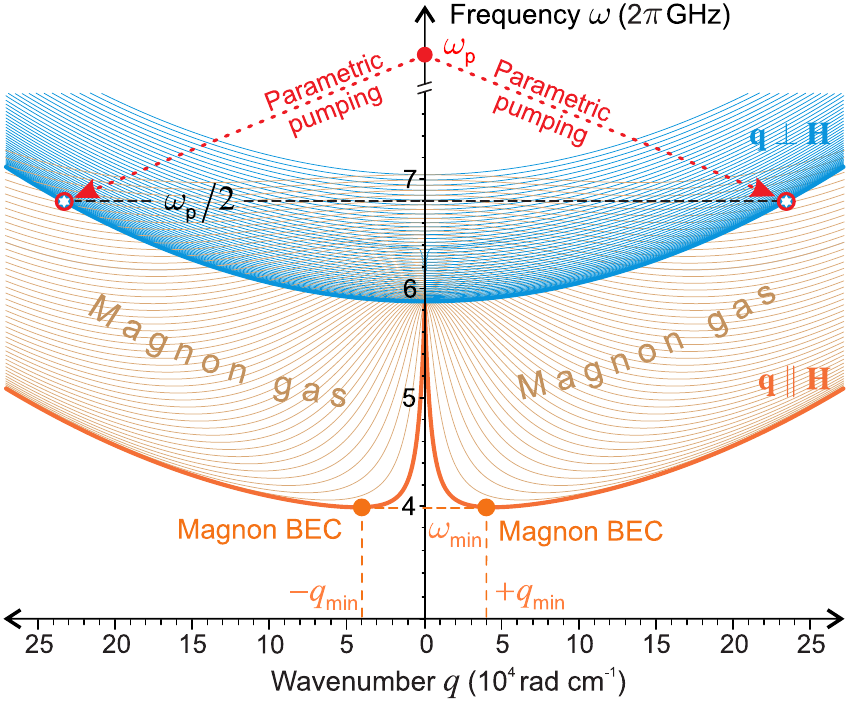}
	\caption{
		\label{f:1} Magnon spectrum of the first 48 thickness modes in 5.6-$\mu$m-thick YIG film magnetized in plane by a bias magnetic field $H=1400\,$Oe, shown for the wavevector $\B q\parallel \B H$ (lower part of the spectrum, blue curves) and for $ \B q\perp \B H $ (upper part, magenta curves). The red arrow illustrates the magnon injection process by means of parallel parametric pumping. Two orange dots indicate positions of the frequency minimum  $\omega\sb{min}(\pm \B q\sb{min}$) occupied by the BECs of magnons.}     
\end{figure}

The frequency spectrum of magnons in this system, shown in \Fig{f:1}, has two symmetric minima with non-zero frequency and wave-vectors $\omega\sb{min}= \omega (\pm \B q\sb{min})$. The possible BEC has accordingly two components with the wave vectors $\pm \B q\sb{min}$ and the frequency $\omega\sb{min}$. The simplest form of their common wave function is a standing wave:
\begin{equation}\label{SW}
C(\B r,t)=C_0 \cos(\B q\sb {min}\cdot \B r) \exp(-i \omega\sb{min} t)\ .
\end{equation}
In this experiment we create a BEC by microwave radiation of frequency $\omega\sb p\simeq 2\pi \cdot 13.6 \,$GHz that can be considered space homogeneous with wave number $q\sb p \approx 0$. The decay instability of this field with the conservation law \looseness=-1
\begin{equation}\label{DI}
\omega\sb p \Longrightarrow \omega (\B q) + \omega (-\B q)= 2 \omega (\B q) 
\end{equation}
excites ``parametric" magnons with frequency $\omega (\B q)=\omega\sb p/2$ and wave vectors $\pm \B q$. These parametric magnons further interact mainly via $2 \Leftrightarrow 2$ scattering with the conservation laws: 
\begin{equation}\label{2-2}
\omega(\B q_1)+\omega(\B q_2)=\omega(\B q_3)+\omega(\B q_4)\,,\  \B q_1 + \B q_2  = \B q_3 + \B q_4  \,,
\end{equation}
that preserves the total number af magnons and their energy. It is known from the theory of weak wave turbulence\,\cite{ZLF} (see also \Ref{AFM}) that the scattering process\,\eqref{2-2} results mostly in a flux of energy towards large $q$, which leads to a nonessential accumulation of energy at large $q$, and to a flux of magnons toward small $q$. This in turn results in an accumulation of magnons near the bottom of the frequency spectrum $\omega\sb{min}$. The same  $2 \Leftrightarrow 2$ processes\,\eqref{2-2} lead to effective thermalization of the bottom magnons during some time $\tau\sb{th} \lesssim 50-70$\,ns and the subsequent creation of the BEC state \cite{Serga2014,Clausen2015}. 

\begin{figure}[t]
	\includegraphics[width=0.98\columnwidth]{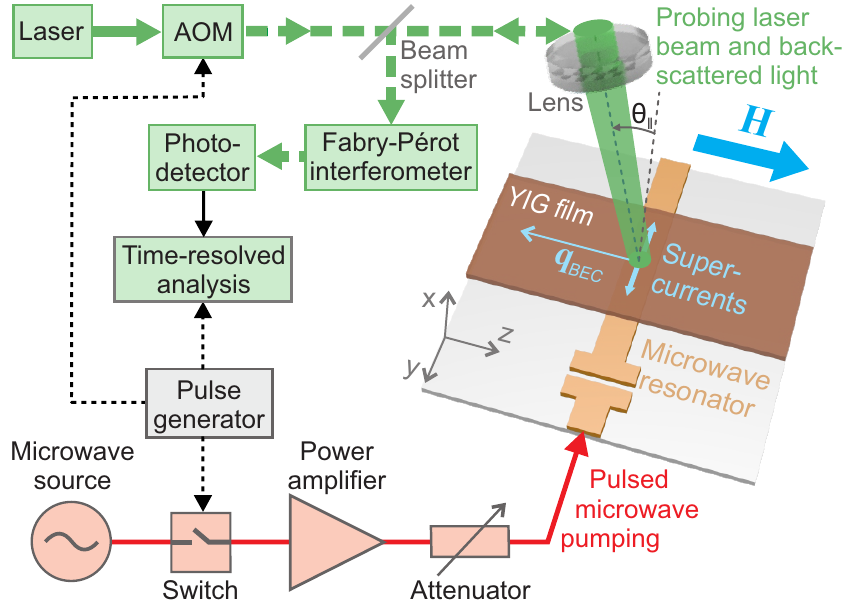}
	\caption{
		\label{f:2} Experimental set-up. The lower part of the figure shows the microwave circuit, consisting of a microwave source, a switch and an amplifier. This circuit drives a microstrip resonator, which is placed below the in-plane magnetized YIG film. Light from a solid-state laser ($\lambda = 532$\,nm) is chopped by an acousto-optic modulator (AOM) and guided to the YIG film. There it is scattered inelastically from magnons, and the frequency-shifted component of the scattered light is selected by the tandem Fabry-P\'{e}rot interferometer, detected, and analyzed in time. }
\end{figure}

The described processes that lead to the creation of BEC, are an experimental manifestation of the space-time crystal (STC): a system, driven away from thermodynamic equilibrium by a space-homogeneous,  time-periodic  (with frequency $\omega\sb p$) pumping field, spontaneously chooses a space-time periodic state\,\eqref{SW} with the frequency $\omega\sb{min}$ and non-zero wavevectors $\pm \B q\sb{min}$. Importantly, the parameters $\omega\sb{min}$ and $\pm \B q\sb{min}$ are fully determined by intrinsic interactions in the system and are independent of the pumping frequency in a wide range of its values. By varying the strength and direction of the external time-independent homogeneous magnetic field $\B H$, the temperature $T$ and the thickness of the YIG film, we can change the magnon spectrum  $\omega(\B q)$ and consequently  $\omega\sb{min}$ and $\pm \B q\sb{min}$ independently of $\omega\sb p$. Note that the lifetime $\tau\Sb{BEC}$ of the condensate  is much longer than $\tau\sb{th}$, enabling the observation of the magnon BEC state and the study of related effects, such as  magnon supercurrent\,\cite{Bozhko2016} and Nambu-Goldstone modes---the Bogolyubov second sound\,\cite{SS2018}. All these meet the presently accepted criteria of a space-time crystal, i.e the spontaneous symmetry breaking in time and in space, manifested by long-range order and soft modes\,\cite{STC-4} (in our case the Bogolyubov second sound\,\cite{SS2018}).

\begin{figure}[t]
	\includegraphics[width=1.\columnwidth]{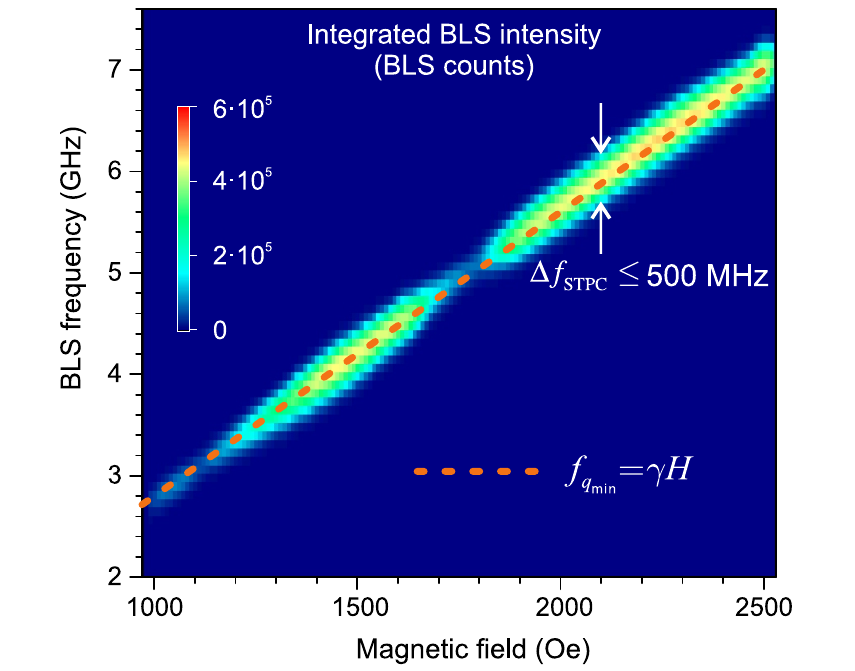}
	\caption{
		\label{f:4}BLS intensity (color code) measured for $q=q\sb{min}$ as a function of the magnon frequency $f$ and of the bias magnetic field $H$. Film thickness $5.6\,\mu\text{m}$, pumping power $40\,\text{W}$, pumping frequency $14\,\text{GHz}$, pumping pulse duration $1\,\mu\text{s}$, pumping period $200\,\mu\text{s}$. The dashed line represents the analytical dependence of the frequency of the spin wave spectrum bottom on $H$: $f_{q\sb{min}}(H)=\gamma H$, where the gyromagnetic ratio $\gamma = 2.8$\,MHz/Oe. 
	}
\end{figure}

The BEC is created from the gaseous incoherent magnons, that accumulate in a relatively narrow frequency band $\Delta f_\mathrm{_{STPC}}$ near the bottom of the spectrum. To keep in line with the crystal analogy, we will refer to this state as a space-time-polycrystal (STPC). In our measurements, the autocorrelation time of these magnons ($ 1/\Delta f_\mathrm{_{STPC}} \geq 2$\,ns, see Fig.\,\ref{f:4}) significantly exceeds the wave period $2\pi/\omega\sb{min} \approx 0.15 - 0.3$\,ns, similarly to the autocorrelation length in polycrystals that spans many unit cell sizes. 

In our experiments, the magnon BEC in the room-temperature YIG films was detected by means of pulsed Brillouin light scattering (BLS) spectroscopy. Here the focused laser beam acts both as a probe of the magnon density and as a heating source, which induces a thermal gradient across the probing light spot. The temperature in the spot, and thus the value of thermal gradient, was controlled by the duration of a probing laser pulse. The thermal gradient locally changes the saturation magnetization and induces a frequency shift between different parts of the magnon condensate \cite{Vogel2015}. Consequently, a phase gradient in the BEC wavefunction is gradually created and a magnon supercurrent \cite{Bozhko2016,Kreil2018}, flowing out of the hot region of the focal spot is excited. Such a process reduces the number of magnons in the heated area and results in the disappearance of the condensate and in the subsequent disappearance of the supercurrent. The conventional relaxation dynamics of the magnons is then recovered. More details about our experimental techniques one finds in a sketch of the experimental setup shown in Fig.\,\ref{f:2}, in Ref.\,\cite{Bozhko2016} and in the supplementary material. 

%\noindent\textbf{\textit{ Experimental results and their discussion}}.

\begin{figure} 
	\includegraphics[width=0.98\columnwidth]{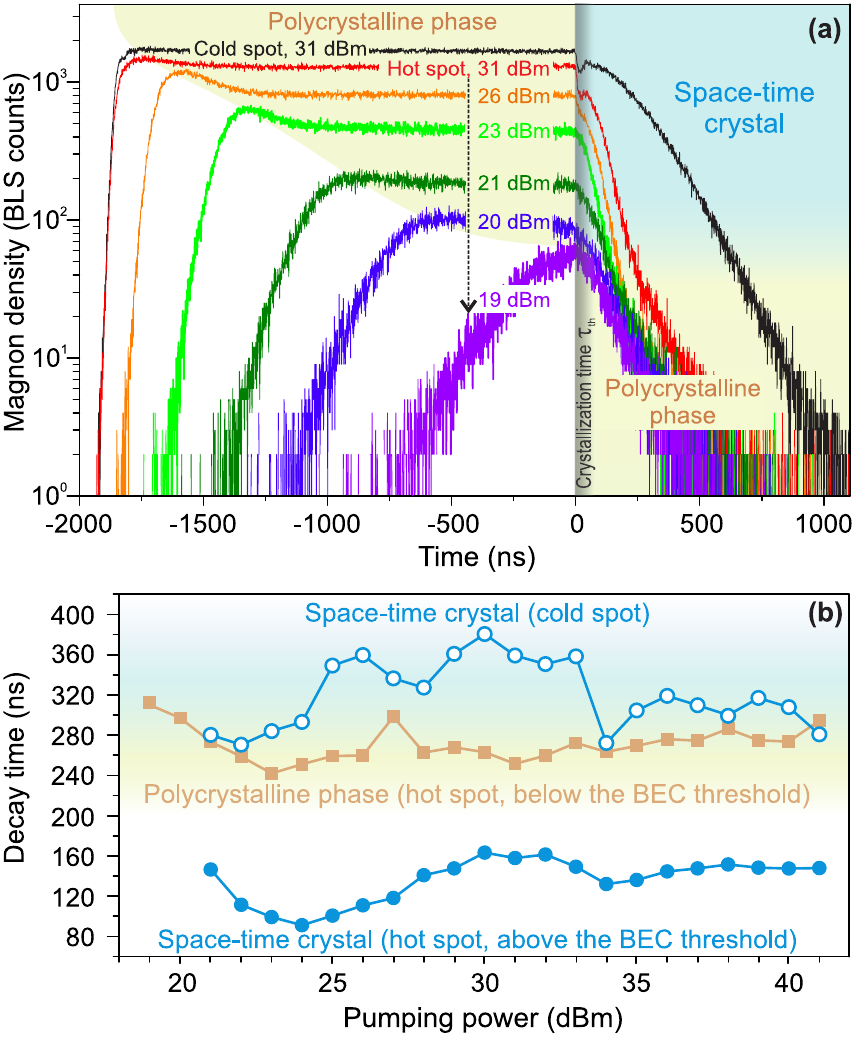}
	\caption{
		\label{f:3} Transition from the polycrystalline magnon phase to the space-time crystal phase and back.  
		(a) Temporal dynamics of the measured magnon density for a few pumping power values $P_\mathrm{pump}$ at different temperatures of the probing spot. The bias magnetic field $H=1400$\,Oe. The top BLS waveform measured for $P_\mathrm{pump}=31$\,dBm corresponds to the case of the weakly heated YIG sample (duration of the probing laser pulse $\tau_{\mathrm{L}}=6\,\mu$s) and, therefore, is not affected by a supercurrent magnon outflow. In all other cases, the non-uniform heating of the YIG sample ($\tau_{\mathrm{L}}=80\,\mu$s) creates a magnon supercurrent flowing out from the heated area resulting in a higher decay rate of the magnons in the BEC phase. This effective decay rate falls with the pumping power. Below a critical magnon density $N_{\mathrm{cr}}$, characterizing the transition from the space-time crystal phase back to the polycrystalline phase, the decay rate is approximately the same for all cases. 
		(b) The decay times $\tau_{\mathrm{dec}}$ of the space-time crystal phase (open and solid circles) and polycrystalline phase (squares) as functions of the pumping power $P_{\mathrm{pump}}$. The space-crystal phase does not exist at pumping powers below 21\,dBm.
	     \looseness=-1}
\end{figure}

We demonstrate here how to change all three parameters of the STC, Eq.\,\eqref{SW}: the  BEC magnon density  $|C_0|^2=N\Sb{BEC}$, the frequency $\omega\sb{min}$ and the wave number $\pm q\sb{min}$. The STC lifetime can be controlled as well. The most interesting information may be obtained by varying $N\Sb{BEC}$. We succeeded to change $N\Sb{BEC}$ by more than an order of magnitude by tuning the power of the pumping field. The measured BLS intensity is shown in Fig.\,\ref{f:3}a as a function of time for selected pumping powers $P_{\mathrm{pump}}$ and two probing laser pulse durations $\tau\Sb{L}$. The pumping pulse acts during the time interval from $-2000\,$ns to 0\,ns. Clearly, a decrease in the pumping power from $P_\mathrm{{pump}}=31$\,dBm to 19\,dBm and a consequent reduction in the number of parametric magnons, which are injected at $\omega(\B q)=\omega\sb p/2$, leads to  a weakening of $2 \Leftrightarrow 2$ magnon scattering and, thus, to an increasing delay in the appearance of these magnons near the bottom of the energy spectrum, as observed by BLS. The density of the bottom magnons,  proportional to the intensity of the measured BLS signal, decreases as well (see the yellow shaded area in Fig.~\ref{f:3}a, labeled ``Polycrystalline phase''). \looseness=-1

After the pumping pulse is switched off (for $t> 0\,$ns), the magnons condense in the energetic minimum of the spectrum, creating the STC. In case of strong heating ($\tau\Sb{L}=80\,\mu$s), this process results in the appearance of a magnon supercurrent, which only involves the condensed and therefore coherent magnons. This outflow of magnons (blue shaded area in Fig.\,\ref{f:3}a labeled \textit{space-time crystal}) results in a higher decay rate of the magnon density in the laser focal point. This effective decay rate, which is influenced by the inherent damping of both coherent and incoherent magnons to the phonon bath and by the supercurrent-related leakage of the magnon BEC, is strongly dependent on the pumping power. This dependence  stems from the fact, that a lower pumping power leads to a reduced magnon density and therefore to a smaller fraction of BEC magnons. Below a certain threshold density $N_{\mathrm{cr}}$, when the majority of condensed magnons are flown out of the measured region of the BEC, the observed decay rate approaches the same value for all different pumping powers. This decay rate corresponds to the inherent decay rate of a narrow package of the remaining polycrystalline magnon phase.
 
It is worth noting that the same decay rate is observed during the entire decay time when heating of the YIG sample can be neglected, and therefore there is no supercurrent that takes away the coherent magnons. For example, the black top waveform in Fig.\,\ref{f:3}a was measured at shorter heating times $\tau_{\mathrm{L}}=6\,\mu$s. The pumping power is the same as for the red waveform  (hot spot, $P_\mathrm{pump}=31$\,dBm), therefore it corresponds to a well-formed BEC. However, it is not possible to distinguish between the BEC and the incoherent magnons via the decay rate measurements in this case. The latter fact contradicts a previous interpretation of similar dynamics of the magnon BEC and the incoherent magnon phase in Ref.\,\cite{Demidov2008} as being a result of the sensitivity of the BLS technique to the degree of coherence of the scattering magnons. Furthermore, two different lifetimes of BEC observed at the same pumping power prove our ability to control the lifetime of the magnon BEC by a thermal gradient.
\looseness=-1

Thereby, the density (Fig.\,\ref{f:3}a) and the lifetime (Fig.\,\ref{f:3}b) of the magnon space-time crystal are tunable by the parametric pumping power and by the proper adjustment of a spatial temperature profile. \looseness=-1
 
The time periodicity $1/\omega\sb{min}$ of the STC can be easily changed by variation of the bias magnetic field. Figure\,\ref{f:4} shows the BLS intensity from the bottom of the magnon spectrum ($\B q=\B q\sb{min}$) as a function of the frequency and the magnetic field. The color coded intensity of the BLS reflects the efficiency of the parametric magnon transfer to the bottom of the frequency spectrum during the pumping pulse \cite{Kreil2018}. The dependence $f_{q\sb{min}}(H)$ is well described by the analytical dependence $\omega\sb{min}= 2\pi f_{q\sb{min}}(H) =2\pi \gamma H$, where $\gamma$ is the gyromagnetic ratio. 
    
\begin{figure} [t]
	\includegraphics[width=1 \columnwidth]{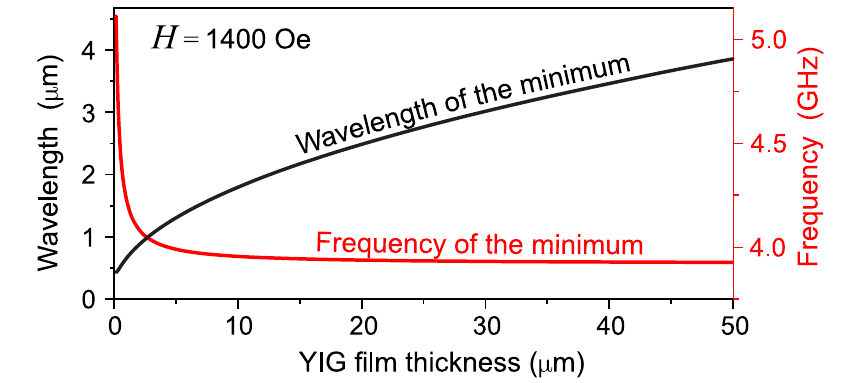}
	\caption{
		\label{f:5} Theoretical  dependencies of the  energy minimum  wavelength $\lambda\sb{min}$ (black line) and frequency $\omega\sb{min}$ (red line) on the YIG film  thickness for $T=300\,$K and $H=1400\,\text{Oe}$.
	}
\end{figure}

The spatial periodicity of the STC can be changed in a wide range from $\simeq 0.5\,\mu$m  to $\simeq 4\,\mu$m by a proper choice of the YIG film thickness, see \Fig{f:5}. Note that, except for very thin films, the $\omega\sb{min}$ is insensitive to the film thickness.

%\noindent \textbf{ \textit{Towards Floquet nonlinear wave physics}}. 
 
The tunable magnon space-time crystal, realized by a periodically driven room-temperature YIG film, represents an example of a nonlinear Floquet system and  therefore serves as a bridge between magnonics and classical  nonlinear wave physics from one side and the Floquet time-crystal description of the periodically driven systems from another. Joining these two perspectives may give  birth to a new field of physical research: ``Floquet (or periodically driven) nonlinear wave physics". 
 
The advantage of a macroscopic system that may be studied at room temperature as compared to small samples at milli-Kelvin temperatures, is obvious. Moreover, strong nonlinearity, non-reciprocity, topology, local manipulation via external electric and magnetic fields and sample patterning, available in a magnonic system, combined with tunability and space-, time-, wave-vector- and frequency-resolved measurements using BLS, makes the suggested system a good experimental basis for the newly proposed field. On the other hand, concepts discussed in the framework of the Floquet systems such as quasi-energy, umklapp scattering, forbidden bands in quasi-momentum space, once applied to magnon space-time crystals, may give new insight into the rich physics of this system, creating new physical ideas and paving a way to new engineering applications. 
    
% \subsection*{Acknowledgments} 
Financial support by the European Research Council within the Advanced Grant 694709 ``SuperMagnonics'' and by Deutsche Forschungsgemeinschaft (DFG) within the Transregional Collaborative Research Center SFB/TR49 ``Condensed Matter Systems with Variable Many-Body Interaction'' as well as by the DFG Project INST 248/178-1 is gratefully acknowledged.

\newpage
%~\\
%  \centerline{\white{***}}
  
\noindent
  \textbf{{Supplemental material: Experimental technique}}  \\
Figure~\ref{f:2} provides a sketch of the experimental setup. The sample is placed between the poles of an electromagnet, which creates a homogeneous magnetizing field $\B H$ lying in the plane of YIG film. In order to reach a high enough density of the magnon gas to form a BEC phase, a rather strong microwave pulse with the peak power $P_\mathrm{max}=41$\,dBm is applied to the half-wave microstrip resonator at a carrier frequency $f_\mathrm{pump}=13.6$\,GHz. The resonator creates an Oersted field $\B q(t) \parallel \B H$ in the YIG film to excite magnons by means of parallel parametric pumping \cite{MOW, NSW}. Additionally, a variable attenuator is implemented in the microwave circuit to allow for power reduction of the amplified microwave pulse to a lower value $P_\mathrm{pump} \leq {P_\mathrm{max}}$. The microwave pulse duration is kept constant at $\tau_{\mathrm{p}}=2\,\mu$s with a repetition rate of $f_{\mathrm{rep}}=1$\,kHz to ensure that microwave heating effects are negligible.

Both the detection of the excited magnons and the heating of the YIG film were made by using the probing laser beam with a power of 30\,mW. The beam is chopped by an acusto-optic modulator (AOM) to control the energy input into the YIG sample. A duration of the laser pulse of $\tau_{\mathrm{L}}=80\,\mu$s is used for local heating, while the pulse duration $\tau_{\mathrm{L}}=6\,\mu$s is used to avoid the heating of the probing point. The probing laser pulse is synchronized with the microwave pumping and has the same repetition rate $f_\mathrm{{rep}}$. The laser pulses of both durations are switched on before the application of the microwave pulse and are switched off 3\,$\mu$s after its end.
From the sample backscattered laser light is collected and sent to a Tandem-Fabry-P\'erot interferometer for frequency and time-of-flight analysis with a frequency and time resolution of approximately 100\,MHz and 1\,ns.

In order to selectively detect only the magnons condensed in the lowest energy state of the magnonic system with a wave number $q_{\parallel}\approx 4.5$\,rad~$\mu$m$^{-1}$ ($\B q_\parallel \parallel \B H$) and a frequency $f_{\mathrm{min}}= 4$\,GHz (see Fig.~\ref{f:2}, the angle of incidence $\theta_{\parallel}$ of the probing laser beam has to be chosen accordingly.
The condition to detect a magnon with a specific wavevector $\mathbf{q}_{\mathrm{_{SW}}}$ lying in a film plane is $\B q_{\mathrm{_{SW}}}=2\B q_\mathrm{light}\sin(\vartheta)$.
Therefore the angle of the incident light has been chosen to be $\theta_{\parallel}=12^\circ$.

\end{document}